\documentclass[onecolumn]{revtex4}

\usepackage{graphicx}
\usepackage{dcolumn}
\usepackage{bm}

\input epsf

\begin{document}

\title{\Large Role of generalized Ricci dark energy on chameleon field in the emergent universe}
\author{\bf  Surajit
Chattopadhyay$^1$$^,$$^2$\footnote{surajit$_{_{-}}2008$@yahoo.co.in,
surajit.chattopadhyay@pcmt-india.net} and Ujjal
Debnath$^2$\footnote{ujjaldebnath@yahoo.com ,
ujjal@iucaa.ernet.in}}

\affiliation{$^1$Department of Computer Application (Mathematics
Section), Pailan College of Management and Technology, Bengal
Pailan Park,
Kolkata-700 104, India.\\
$^2$Department of Mathematics, Bengal Engineering and Science
University, Shibpur, Howrah-711 103, India. }

\date{\today}

\begin{abstract}
In this paper, we have considered the generalized Ricci dark
energy (GRDE) and generalized holographic dark energy (GHDE) in
the scenario of emergent universe. Fractional energy density and
deceleration parameters for GRDE were derived under emergent
universe scenario. Also role of GRDE on the Chameleon field in the
emergent universe scenario has been examined. Finally, the
behaviours of the Chameleon scalar field $\phi$, corresponding
potential $V$ and associated function $f$ were investigated in
presence of GRDE.
\end{abstract}


\maketitle

\subsection{\bf\large{Introduction}}

The accelerated expansion of the universe has now been well
documented in the literature [1] and is strongly confirmed by the
cosmic microwave background radiation (CMBR) [2] and Sloan Digital
Sky Survey (SDSS)[3]. Another way of presenting this kinematic
property of the universe is to postulate the existence of a new
entity — dark energy (DE) that has been reviewed in several
literatures [4]. Although observationally well-established, no
single theoretical model provides an entirely compelling framework
within which cosmic acceleration or DE can be understood. The
simplest candidate of dark energy is the vacuum energy density or
cosmological constant $\Lambda$, whose energy density remains
constant with time $\rho_{\Lambda} = \Lambda/8\pi G$ and whose
equation of motion is also fixed,
$w=p_{\Lambda}/\rho_{\Lambda}=-1$  ($p_{\Lambda}$ is the pressure)
during the evolution of the universe. The resulting cosmological
model, $\Lambda$CDM, consists of a mixture of vacuum energy and
cold dark matter [5]. Another possibility is QCDM cosmologies
based on a mixture of cold dark matter and quintessence
$(-1<w\leq0)$ [6]. Numerous other candidates for dark energy have
also been proposed in the literature, such as an evolving
canonical scalar field [7] usually referred to as quintessence,
the phantom energy [8] with an equation of state smaller than $-1$
violating the weak energy condition, the quintom energy [9] with
an equation of state evolving across $-1$, and so forth.
\\\\
 In recent times, an interesting attempt for probing the nature of
dark energy within the framework of quantum gravity is the
so-called ''holographic dark energy'' (HDE) proposal [10]. The
holographic principle is an important result of the recent
research for exploring the quantum gravity [11]. This principle is
enlightened by investigations of the quantum property of black
holes [e.g. 12]. In a quantum gravity system, the conventional
local quantum field theory will break down because it contains too
many degrees of freedom that will lead to the formation of a black
hole breaking the effectiveness of the quantum field theory [11].
The HDE model is constructed by considering the holographic
principle and some features of the theory of quantum gravity. The
HDE constructed in light of the holographic principle possesses
some significant features of an underlying theory of dark energy
[13]. According to the holographic principle, the number of
degrees of freedom in a bounded system should be finite and has
relations with the area of its boundary. The energy density of HDE
is given by $\rho_{\Lambda}=3 c^{2}M_{pl}^{2}L^{-2}$ [14], where
$c$ is a numerical constant and $M_{pl}=1/\sqrt{8 \pi G}$ is the
reduced Plank mass. If we take $L$ as the size of the current
Universe, for instance, the Hubble radius $H^{-1}$, then the dark
energy density will be close to the observational result. This
reflects that means that there is duality between UV cut-off and
IR cut-off. The UV cut-off is related to the vacuum energy, and
the IR cut-off is related to the large scale of the universe, for
example the Hubble horizon, event horizon or particle horizon [14,
15].
\\\\
 Inspired by the holographic principle, Gao et al.[16] took the Ricci scalar
as the IR cut-off and named it the Ricci dark energy (RDE), in
which they take the Ricci scalar $R$ as the IR cutoff. With proper
choice of parameters the equation of state crosses $-1$, so it is
a 'quintom' [13]. The Ricci scalar of FRW universe is given by
$R=-6\left(\dot{H} +2H^{2}+\frac{k}{a^{2}}\right)$, where $H$ is
the Hubble parameter, $a$ is the scale factor and $k$ is the
curvature. It has been found that this model does not only avoid
the causality problem and is phenomenologically viable, but also
naturally solves the coincidence problem [17]. Feng and Li [18]
discussed viscous RDE model by assuming that there is bulk
viscosity in the linear barotropic fluid and the RDE. Lepe and
Pena [19] studied the dark energy problem by adopting a
holographic model by postulating an energy density $\rho\sim R$,
where R is the Ricci scalar curvature and showed that the equation
of state for the dark energy exhibits a cross through the $-1$
barrier. Feng [13] adopted a correspondence between the RDE model
and a scalar field dark energy and reconstructed quintom from RDE.
In another work, Feng [20] regarded the $f(R)$ theory as an
effective description for the acceleration of the universe and
reconstruct the function $f(R)$ from the RDE. State finder
diagnostics of RDE was investigated by Feng [21].
\\\\
The possibilities of an emergent universe [22] have been studied
recently in a number of papers in which one looks for a universe
which is ever-existing and large enough so that the spacetime may
be treated as classical entities. In these models, the universe in
the infinite past is in an almost static state but it eventually
evolves into an inflationary stage [23]. An emergent universe
model can be defined as a singularity free universe which is ever
existing with an almost static nature in the infinite past
$(t\rightarrow-\infty)$ and then evolves into an inflationary
stage [24]. Debnath [25] investigated the emergent universe
scenario under the situation that the universe is filled with
normal matter and phantom field or tachyonic field and derived
various conditions for the possibility of the emergent universe in
the flat, closed and open universes. Mukherjee et al [23] showed
that the emergent universe scenarios are not isolated solutions
and they may occur for different combinations of radiation and
matter. Campo et al [26] studied the emergent universe model in
the context of a self-interacting Jordan-Brans-Dicke theory and
showed that the model presents a stable past eternal static
solution which eventually enters a phase where the stability of
this solution is broken leading to an inflationary period.
Mukherjee et al [23] summarized the features of emergent universe
as:
\begin{enumerate}
    \item the universe is almost static at the finite past $(t\rightarrow-\infty)$ and isotropic, homogeneous at large
    scales;
    \item it is ever existing and there is no timelike
    singularity;
    \item the universe is always large enough so that the classical
description of space-time is adequate;
    \item the universe may contain exotic matter so that the energy conditions
may be violated;
    \item the universe is accelerating as suggested by recent measurements of distances of
high redshift type Ia supernovae.
\end{enumerate}

A generalized model has been designed by Xu et al [27] to included
HDE and RDE by introducing a new parameter which balances
holographic and Ricci dark energy model. Xu et al [27] considered
the energy densities for generalized holographic dark energy
(GHDE) and generalized Ricci dark energy (GRDE) as

\begin{equation}
\rho_{GH}=3c^{2}M_{pl}^{2}~f\left(\frac{R}{H^{2}}\right)H^{2}
\end{equation}

\begin{equation}
\rho_{GR}=3c^{2}M_{pl}^{2}~g\left(\frac{H^{2}}{R}\right)R
\end{equation}

where $f(x)$ and $g(y)$ are positive defined functions of the
dimensionless variables $x=R/H^{2}$ and $y=H^{2}/R$ respectively.
Xu et al (2009) had chosen

\begin{equation}
f\left(\frac{R}{H^{2}}\right)=1-\epsilon
\left(1-\frac{R}{H^{2}}\right)
\end{equation}

\begin{equation}
g\left(\frac{H^{2}}{R}\right)=1-\eta
\left(1-\frac{H^{2}}{R}\right)
\end{equation}

where $\epsilon$ and $\eta$ are parameters. When $\epsilon=0 (\eta
=1)$ or $\epsilon=1 (\eta =0)$, the generalized energy density
becomes the holographic (Ricci) and Ricci (holographic) dark
energy density respectively. If $\epsilon=1-\eta$, then GHDE and
GRDE are equivalent.
\\\\
One possibility to explain the current accelerated expansion of
the universe may be related with the presence of cosmologically
evolving scalar whose mass depends on the local matter density
(chameleon cosmology) [28].In the flat homogeneous universe the
action of chameleon field [29] is given by [30]

\begin{equation}
S=\int\sqrt{-g}d^{4}x\left[f(\phi)\mathcal{L}+\frac{1}{2}\phi_{,\mu}\phi^{,\mu}+\frac{R}{16\pi
G}-V(\phi)\right]
\end{equation}

where $\phi$ is the Chameleon scalar field and $V(\phi)$ is the
Chameleon potential. Also, $R$ is the Ricci scalar and G is the
Newtonian constant of gravity, $f(\phi)\mathcal{L}$ is the
modified matter Lagrangian and $f(\phi)$ is an analytic function
of $\phi$.  This term brings about the non-minimal interaction
between the cold dark matter and Chameleon field. Reason behind
dubbing the the scalar field as ``Chamelion" is stated in Khoury
and Weltman [29] as: ``We refer to $\phi$ as a `chameleon' field,
since its physical properties, such as its mass, depend
sensitively on the environment. Moreover, in regions of high
density, the chameleon `blends' with its environment and becomes
essentially invisible to searches for equivalence principle
violation and fifth force".
\\\\
In the present paper we have considered emergent scenario of the
universe in presence of GRDE with chameleon field. Following
references [23] and [24] the scale factor has been taken in the
form

\begin{equation}
a=a_{0}\left(\beta+e^{\alpha t}\right)^{n}
\end{equation}

where the constant parameters are restricted as follows [24]:\\
\\
1. $a_{0}>0$ for the scale factor a to be positive, \\
2. $\beta>0$, to avoid any singularity at finite time (big-rip),\\
3. $\alpha>0$ or $n>0$ for expanding model of the universe,\\
4. $\alpha<0$ and $n<0$ implies big bang singularity at
$t=-\infty$.\\
\\\\
\subsection{\bf\large{RDE in emergent universe}}

In this section we are beginning with Ricci dark energy (RDE)
which is characterized by the energy density [16, 27]

\begin{equation}
\rho_{R}=3 c^{2}\left(\dot{H}+2H^{2}+\frac{k}{a^{2}}\right)
\end{equation}

where $c$ is a dimensionless parameter which will determine the
evolution behavior of RDE. When $c^{2}<1/2$, the RDE will exhibit
a quintomlike behavior; i.e., its equation of state will evolve
across the cosmological-constant boundary $w=-1$. Reviews on RDE
are available in papers like [27], [31], [32] and [33]. The
conservation equations for RDE is

\begin{equation}
\dot{\rho}_{R}+3H (\rho_{R}+p_{R})=0
\end{equation}

In a flat universe i.e. $k=0$, we have $\rho_{R}=3
c^{2}(\dot{H}+2H^{2})$. Solving the conservation equation for RDE
in a flat universe one can obtain the pressure as

\begin{equation}
p_{R}=-c^{2}\left(\frac{\ddot{H}}{H}+7\dot{H}+6H^{2}\right)
\end{equation}

With the choice of scale factor for emergent universe we obtain
the equation of state (EOS) parameter $w_{RDE}=p_{RDE}/\rho_{RDE}$
and plot it in figure 1, which shows that the EOS parameter is
staying below $-1$ and gradually tending to $-1$ with passage of
cosmic time. Thus, we find that the RDE is behaving like phantom
in emergent universe.

\begin{figure}
\includegraphics[height=1.8in]{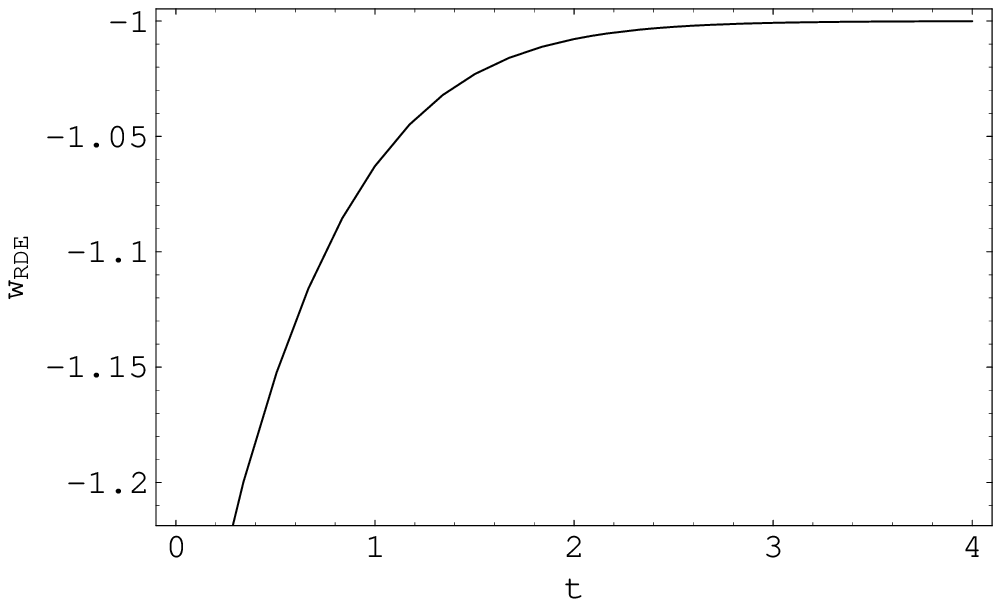}~~~\\
\vspace{1mm} ~~~~~~~~~~~~Fig.1\\
\vspace{6mm} Fig. 1 shows the EOS parameter $w_{RDE}$ for RDE
under emergent universe. The EOS parameter $w_{RDE}<-1$, which
indicates phantom like behaviour.

\vspace{6mm}

\end{figure}

\begin{figure}
\includegraphics[height=1.8in]{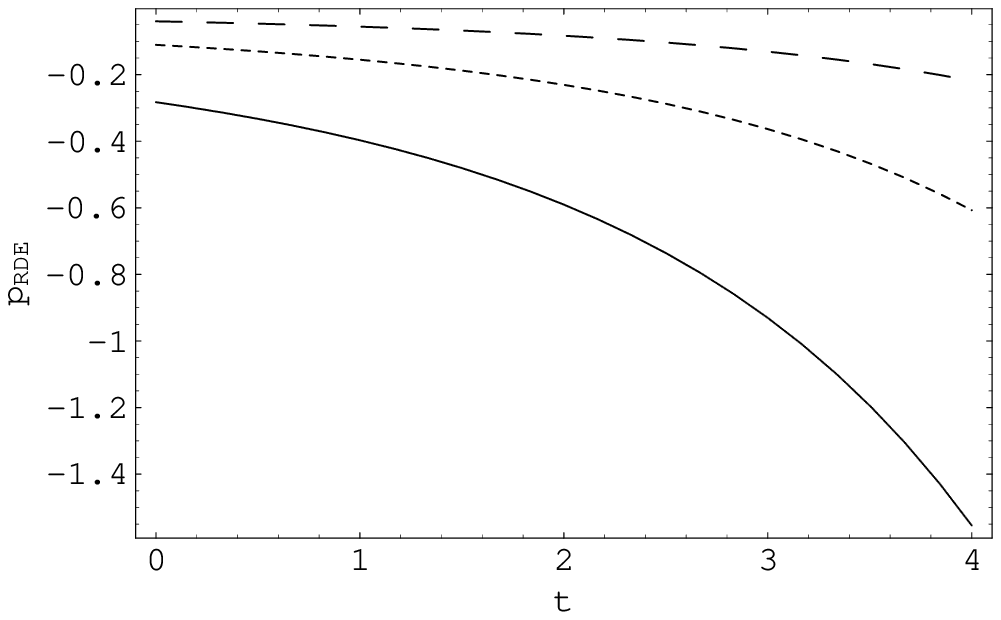}~~~~
\includegraphics[height=1.8in]{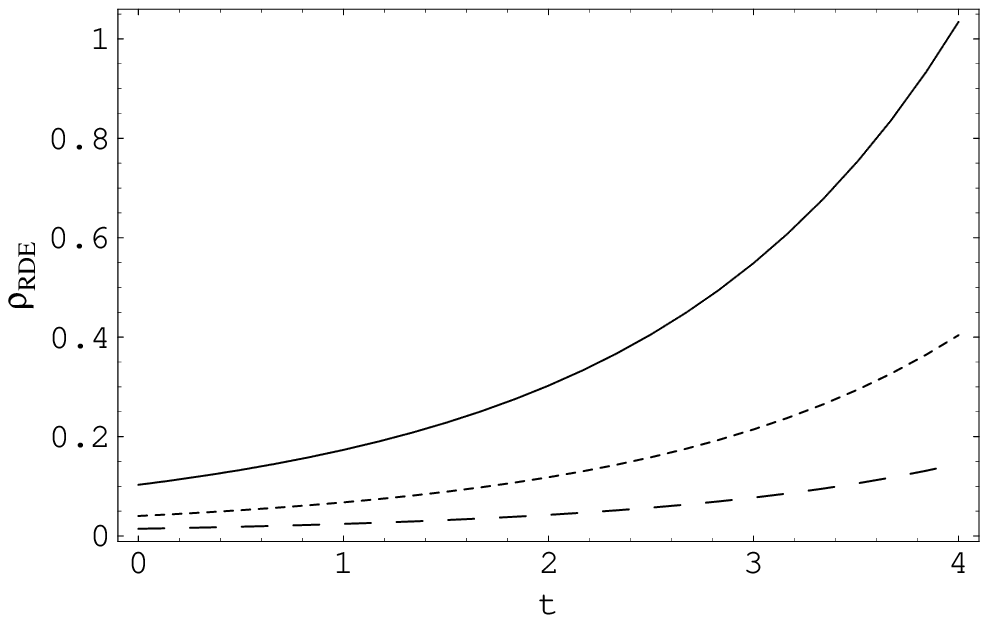}\\
\vspace{1mm}
~~~~~Fig.2~~~~~~~~~~~~~~~~~~~~~~~~~~~~~~~~~~~~~~~~~~~~~~~~~Fig.3\\

\vspace{6mm} ~~~~~~~~~~~~Fig.2 and 3 represent the pressure and energy density of RDE under emergent universe for $c=0.8$ (thick line), $0.5$ (dotted line) and $0.3$ (broken line).\\

\vspace{6mm}

\end{figure}

The pressure and energy density parameters plotted in figures 2
and 3 show that the negative pressure is falling and energy
density is increasing with passage of cosmic time. In the next
section we shall switch over to the generalized situation in
emergent universe scenario.
\\\\

\subsection{\bf\large{GRDE in emergent universe}}
Issues associated with GRDE are already discussed in Section I. In
the case of GRDE, using equations (2) and (4) the Friedman's
equation is

\begin{equation}
H^{2}=\frac{1}{3M_{pl}^{2}}(\rho_{m}+\rho_{GR})=H^{2}\Omega_{m}+c^{2}\left[1-\eta
\left(1-\frac{H^{2}}{R}\right)\right]R
\end{equation}

where $\Omega_{m}=\frac{\rho_{m}}{3M_{pl}^{2}H^{2}}$ is the energy
density for dark matter. In the emergent scenario, the energy
density for GRDE is found to be

\begin{equation}
\begin{array}{c}
  \Omega_{GR}=c^{2}[2-\eta+\frac{H_{0}^{2}a_{0}(e^{t\alpha}+\beta)^{n}(1+c^{2}(-2+\eta))(-2+c^{2}(1+\eta)+2\Omega_{m0})}{c^{2}\left(-2(\frac{(e^{\alpha
t
+\beta})^{-n}}{a_{0}})^{\frac{2(-1+c^{2})}{c^{2}(-1+\eta)}}\Omega_{m0}+H_{0}^{2}(e^{\alpha
t}+\beta)^{n}a_{0}(-2+c^{2}(1+\eta)+2\Omega_{m0})\right)}\\
  +\frac{3(-1+\eta)\Omega_{m0}}{2\Omega_{m0}-H_{0}^{2}(\frac{(e^{\alpha t}+\beta)^{-n}}{a_{0}})^{-1-\frac{2(-1+c^{2})}{c^{2}(-1+\eta)}}}(-2+c^{2}(1+\eta)+2\Omega_{m0})]\\
\end{array}
\end{equation}

Deceleration parameter is computed as

\begin{equation}
q_{GR
}=\frac{1}{1-\eta}-\frac{\Omega_{GR}}{c^{2}(1-\eta)}
\end{equation}

The deceleration and energy density parameters are plotted in
figures 4 and 5.

\begin{figure}
\includegraphics[height=1.8in]{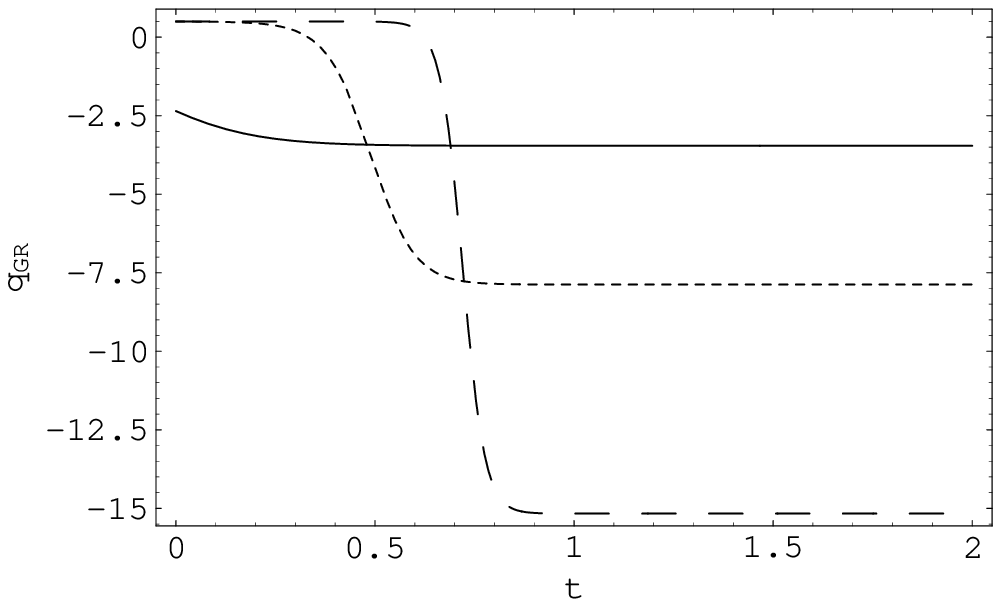}~~~\\
\vspace{1mm} ~~~~~~~~~~~~Fig.4\\
\vspace{6mm} Fig. 4 shows the deceleration parameter $q_{GR}$ for
GRDE under emergent universe. The negative sign of $q_{GR}$ shows
the ever accelerating universe under emergent universe.
$(\eta=1/3; c=0.8, 0.5, 0.3; H_{0}=72)$

\vspace{6mm}

\end{figure}

\begin{figure}
\includegraphics[height=1.8in]{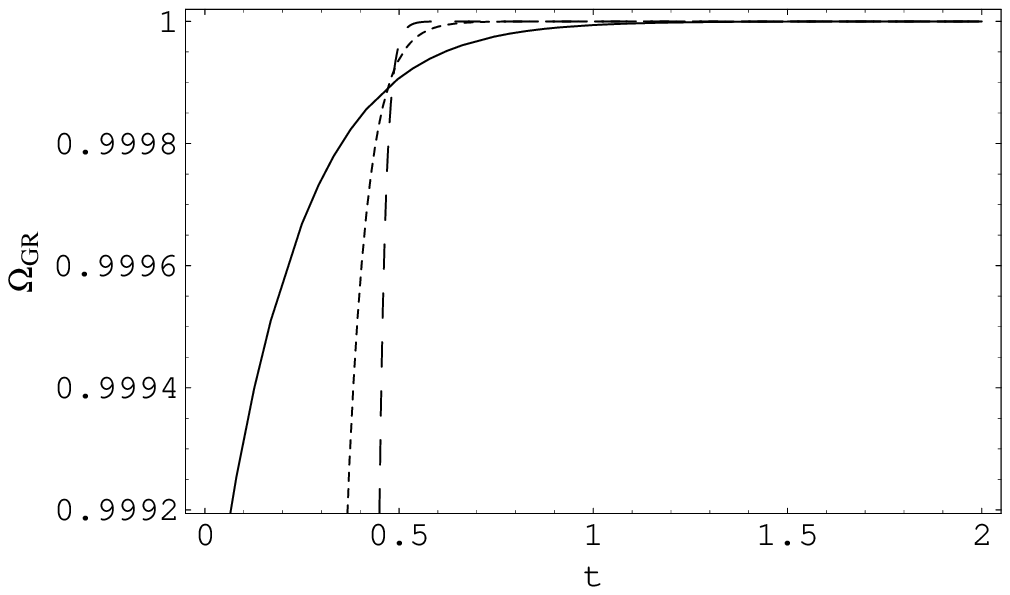}~~~\\
\vspace{1mm} ~~~~~~~~~~~~Fig.5\\
\vspace{6mm} Fig. 5 shows the fractional energy density
$\Omega_{GR}$ of GRDE under emergent universe. The increasing
behaviour of $\Omega_{GR}$ with cosmic time $t$ reflects the dark
energy dominated universe. $(\eta=1/3; c=0.8, 0.5, 0.3; H_{0}=72)$

\vspace{6mm}

\end{figure}

From the negative deceleration parameter throughout the evolution
of the universe the ever accelerating nature of the emergent
universe is reflected. The energy density $\Omega_{GR}$ in figure
5 gradually increases with cosmic time and tends to $1$. This
indicates the dark energy dominated universe. From figure 6, which
depicts the EOS parameter $w_{GR}$, it is apparent that the EOS
parameter is $<-1$. This indicates phantom like behaviour of the
GRDE in the emergent universe scenario. In all these plots we have
taken $\eta=1/3$. We know from reference [27] that if we take
$\epsilon=1-1/3=2/3$ in equation (3) and then use it in equation
(1) then we get a GHDE equivalent to the GRDE discussed above. To
get a different GHDE we we take $\epsilon=3/4$. We obtain the
energy density $\Omega_{GH}$ and deceleration parameter $q_{GH}$
for the GHDE in emergent universe as follows

\begin{equation}
\Omega_{GH}=(1+\epsilon)c^{2}+\frac{-1+c^{2}(1+\epsilon)}{\left(-1-\frac{2(\frac{(e^{t\alpha}+\beta)^{-n}}{a_{0}})^{1+\frac{-2+\frac{2}{c^{2}}}{\epsilon}}\Omega_{m0}}{2+c^{2}(-2+\epsilon)-2\Omega_{m0}}\right)}+\frac{3c^{2}\epsilon\Omega_{m0}}{-2\Omega_{m0}+(\frac{(e^{t\alpha}+\beta)^{-n}}{a_{0}})^{-1+\frac{2-\frac{2}{c^{2}}}{\epsilon}}(-2-c^{2}(-2-c^{2}(-2+\epsilon)+2\Omega_{m0}))}
\end{equation}

\begin{equation}
q_{GH }=\frac{1}{\epsilon}-\frac{\Omega_{GH}}{c^{2}\epsilon}
\end{equation}

\begin{figure}
\includegraphics[height=1.8in]{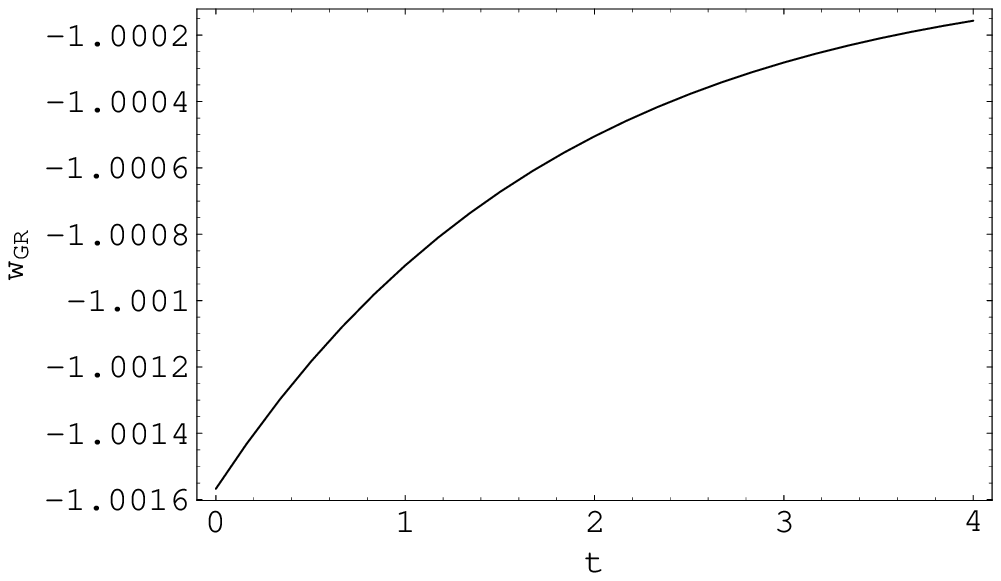}~~~\\
\vspace{1mm} ~~~~~~~~~~~~Fig.6\\
\vspace{6mm} Fig. 6 shows the EOS parameter $w_{GR}$ for GRDE
under emergent universe. The $w_{GR}<-1$ reflects the phantom like
behaviour. $(\eta=1/3; H_{0}=72)$

\vspace{6mm}

\end{figure}

\begin{figure}
\includegraphics[height=1.8in]{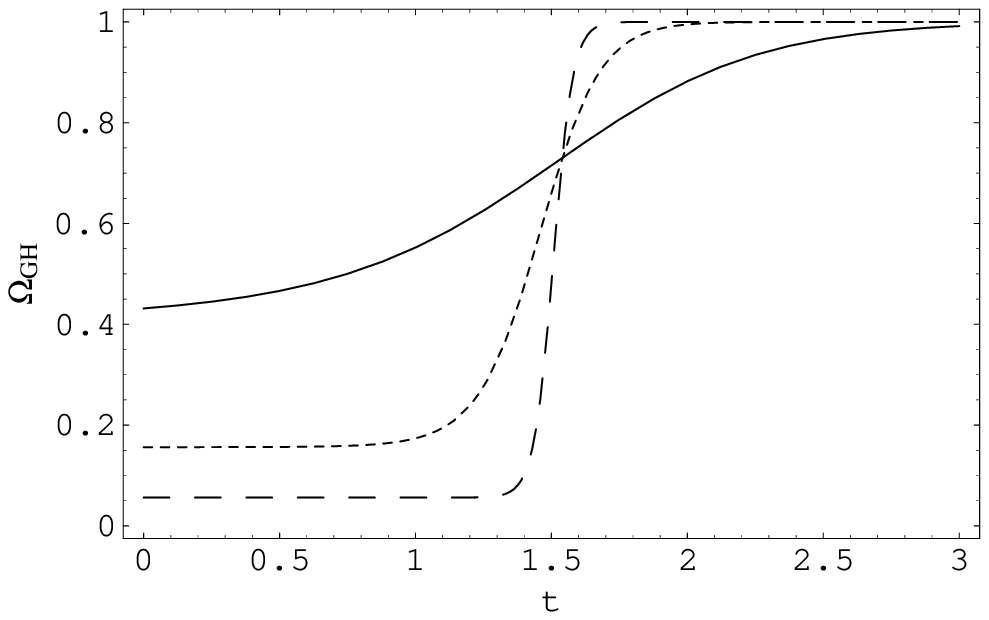}~~~\\
\vspace{1mm} ~~~~~~~~~~~~Fig.7\\
\vspace{6mm} Fig. 7 shows the the fractional energy density
$\Omega_{GH}$ of GHDE under emergent universe. The increasing
behaviour of $\Omega_{GH}$ with cosmic time $t$ reflects the dark
energy dominated universe. $(\epsilon=3/4; c=0.8, 0.5, 0.3;
H_{0}=72)$

\vspace{6mm}

\end{figure}

\begin{figure}
\includegraphics[height=1.8in]{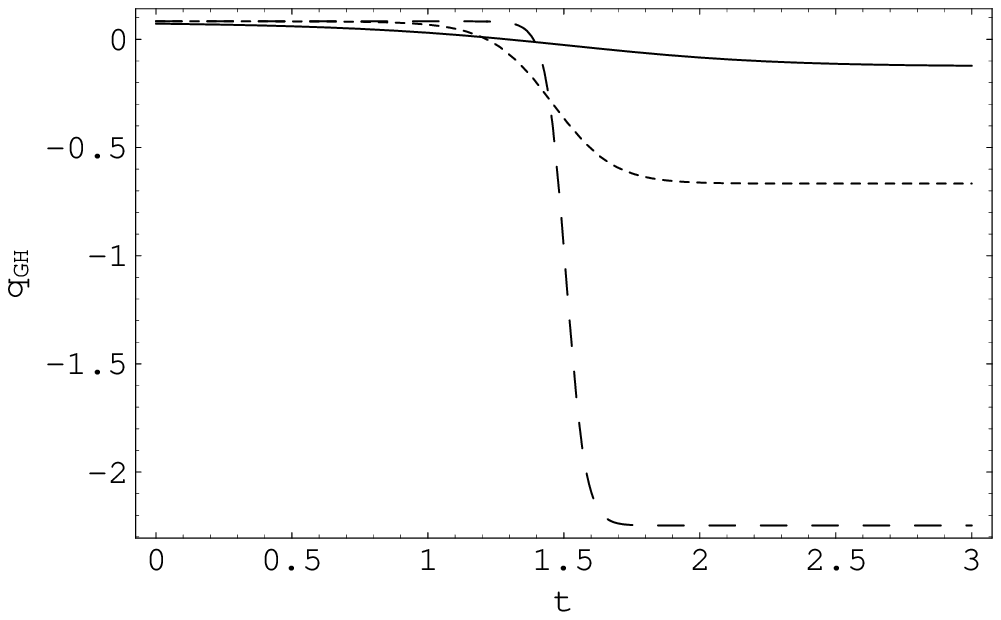}~~~\\
\vspace{1mm} ~~~~~~~~~~~~Fig.8\\
\vspace{6mm} Fig. 8 shows the deceleration parameter $q_{GH}$ for
GHDE under emergent universe. $(\epsilon=3/4; c=0.8, 0.5, 0.3;
H_{0}=72)$

\vspace{6mm}

\end{figure}

\begin{figure}
\includegraphics[height=1.8in]{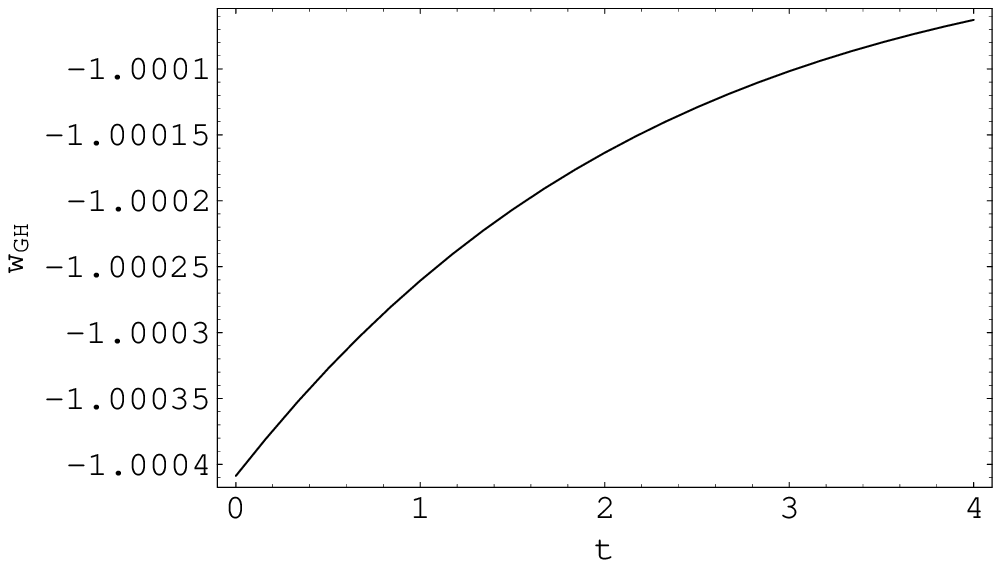}~~~\\
\vspace{1mm} ~~~~~~~~~~~~Fig.9\\
\vspace{6mm} Fig. 9 shows the EOS parameter $w_{GH}$ for GHDE
under emergent universe. The $w_{GH}<-1$ reflects the phantom like
behaviour. $(\epsilon=3/4; H_{0}=72)$

\vspace{6mm}

\end{figure}

The energy density and deceleration parameters are plotted in
figures 7 and 8 respectively. Like GRDE, here also we find a
gradually increasing energy density and a deceleration parameter
which is always negative. This means that we are getting an ever
accelerating universe under emergent scenario with GHDE. The
phantom like behaviour of the EOS parameter of GHDE in emergent
universe scenario is apparent from figure 9. This behaviour is
similar to that of GRDE. In the subsequent section we would
describe the role of GRDE on Chameleon field in the emergent
universe.
\\\\

\subsection{\bf\large{GRDE on chameleon field}}

The variation of action (5) with respect to the metric tensor
components in a FRW cosmology yields

\begin{equation}
3H\dot{\phi}+\ddot{\phi}+\frac{dV}{d\phi}+(\rho+p)\frac{df}{d\phi}=0
\end{equation}

which reduces to

\begin{equation}
3H\dot{\phi}^{2}+\dot{\phi}\ddot{\phi}+\dot{V}+(p+\rho)\dot{f}=0
\end{equation}

which is the wave equation for the chameleon field. Again, the
variation of the same equation with respect to metric tensor
components gives (choosing $8\pi G=1$)

\begin{equation}
3\frac{\dot{a}^{2}}{a^{2}}=\rho f
+\frac{1}{2}\dot{\phi}^{2}+V(\phi)
\end{equation}

\begin{equation}
2\frac{\ddot{a}}{a}+\frac{\dot{a}^{2}}{a^{2}}=-pf-\frac{1}{2}\dot{\phi}^{2}+V(\phi)
\end{equation}

Following reference [34] we choose

\begin{equation}
V=V_{0}\dot{\phi}^{2}
\end{equation}

From the equations (17) and (18) we get

\begin{equation}
\frac{d}{dt}(\rho f)+3H(p+\rho)f=(p+\rho)\dot{f}
\end{equation}

As we are considering the GRDE in the emergent universe, we
replace $\rho$ by
$\rho_{GR}=3c^{2}M_{pl}^{2}(1-\eta(1-\frac{H^{2}}{R}))R$.
Eliminating $\dot{\phi}^{2}$ from the field equations and using
the form of $p$ in equation (20) with
$H=\frac{e^{t\alpha}n\alpha}{e^{t\alpha}+\beta}$ we get (taking
$M_{pl}=1$)

\begin{equation}
\xi_{1}(t)=\frac{e^{t\alpha}n
\alpha^{2}\left[a(3ne^{t\alpha}+2\beta)-3b(6c^{2}\beta(-1+\eta)+e^{t\alpha}n(-1+c^{2}(-12+3\eta)))\right]}{(e^{t\alpha}+\beta)^{2}}
\end{equation}

\begin{equation}
\begin{array}{c}
 \xi_{2}(t)= -\frac{1}{(e^{t\alpha}+\beta)^{4}}[3e^{2t\alpha}n^{2}\alpha^{3}(3b(e^{t\alpha}+\beta)(6c^{2}\beta(-1+\eta)+ne^{t\alpha}(-1+c^{2}(-12+13\eta)))  \\
  +
a(2\beta^{2}(-1+9c^{2}(1+\alpha)(-1+\eta))+3ne^{2t\alpha}(-1+c^{2}(-12+13\eta))\\
  +e^{t\alpha}\beta(-2-3n+3c^{2}(n(1+2\alpha)(-12+13\eta)+6(-1+\eta)(1-\alpha)))))]\\
\end{array}
\end{equation}

where $a=V_{0}+\frac{1}{2}$ and $b=V_{0}-\frac{1}{2}$ and
consequently $f$ is reconstructed as

\begin{equation}
f=f_{0}\left[exp\int\frac{ \xi_{2}(t)}{\xi_{1}(t)}dt\right]
\end{equation}

Now, $f$ is plotted against cosmic time $t$ in figure 10, which
shows that $f$ is increasing with cosmic time $t$.

\begin{figure}
\includegraphics[height=1.8in]{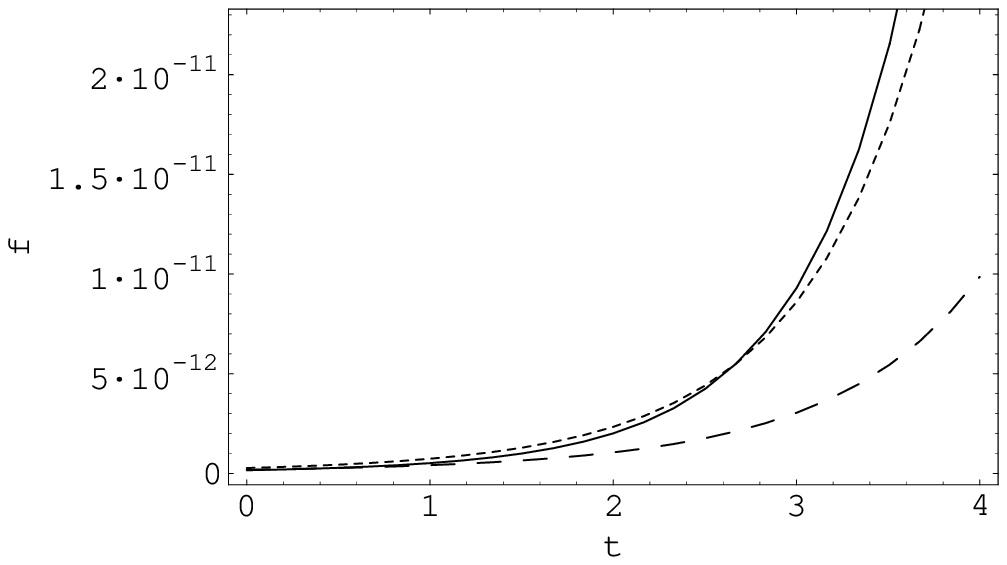}~~~\\
\vspace{1mm} ~~~~~~~~~~~~Fig.10\\
\vspace{6mm} Fig. 10 shows that $f$ increases with cosmic time
$t$. We have taken $\eta=1/3; c=0.8, 0.5, 0.3$.

\vspace{6mm}

\end{figure}

As we know the forms of $H$, $f$ and $\rho_{GR}$, we obtain the
the chameleon scalar field $\phi$ from equation (17) and
consequently the potential $V$ is obtained.

\begin{figure}
\includegraphics[height=1.8in]{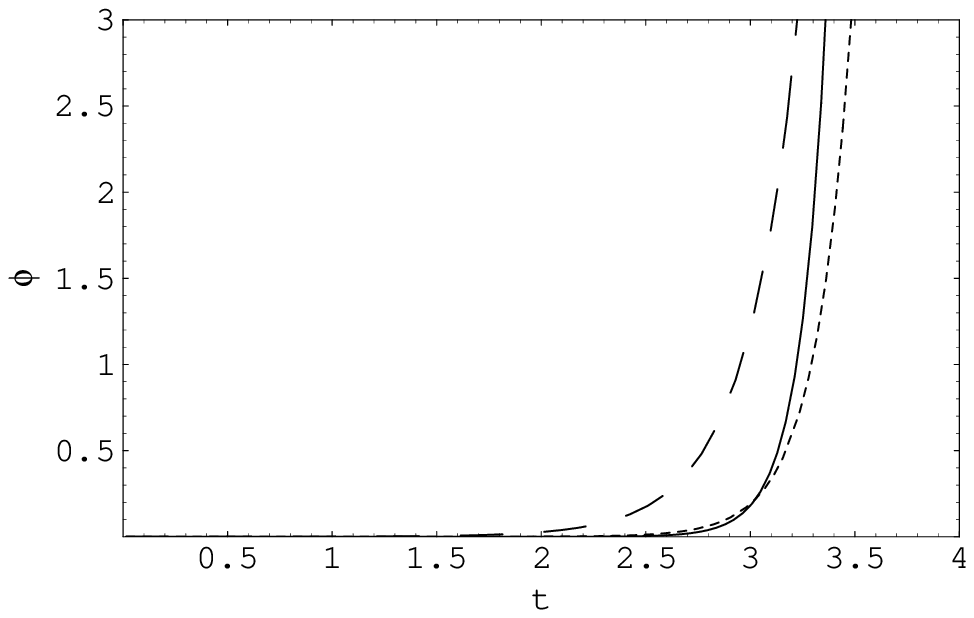}~~~\\
\vspace{1mm} ~~~~~~~~~~~~Fig.11\\
\vspace{6mm} Fig. 11 shows the increasing chameleon scalar field
$\phi$ in presence of GRDE in emergent universe scenario. We have
taken $\eta=1/3; c=0.8, 0.5, 0.3$.

\vspace{6mm}

\end{figure}

\begin{figure}
\includegraphics[height=1.8in]{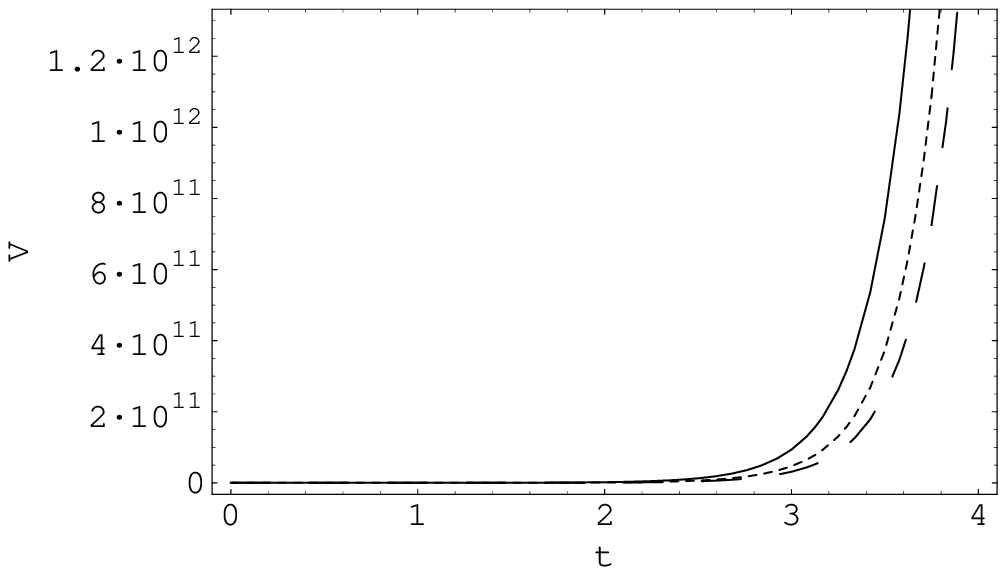}~~~\\
\vspace{1mm} ~~~~~~~~~~~~Fig.12\\
\vspace{6mm} Fig. 12 shows the increasing potential $V$ of the
chameleon field in presence of GRDE in emergent universe scenario.
We have taken $\eta=1/3; c=0.8, 0.5, 0.3$.

\vspace{6mm}

\end{figure}

\begin{figure}
\includegraphics[height=1.8in]{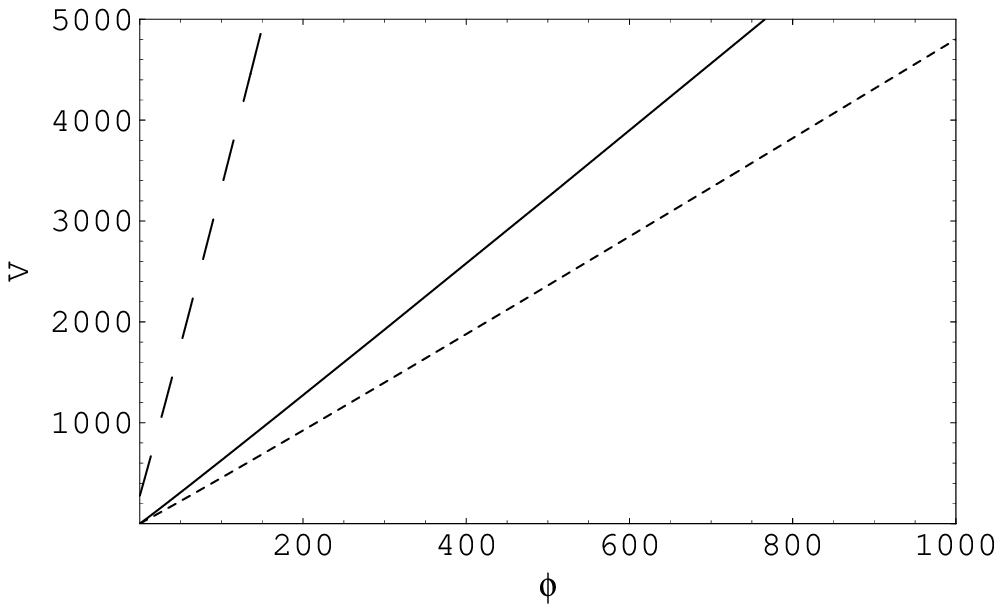}~~~~
\includegraphics[height=1.8in]{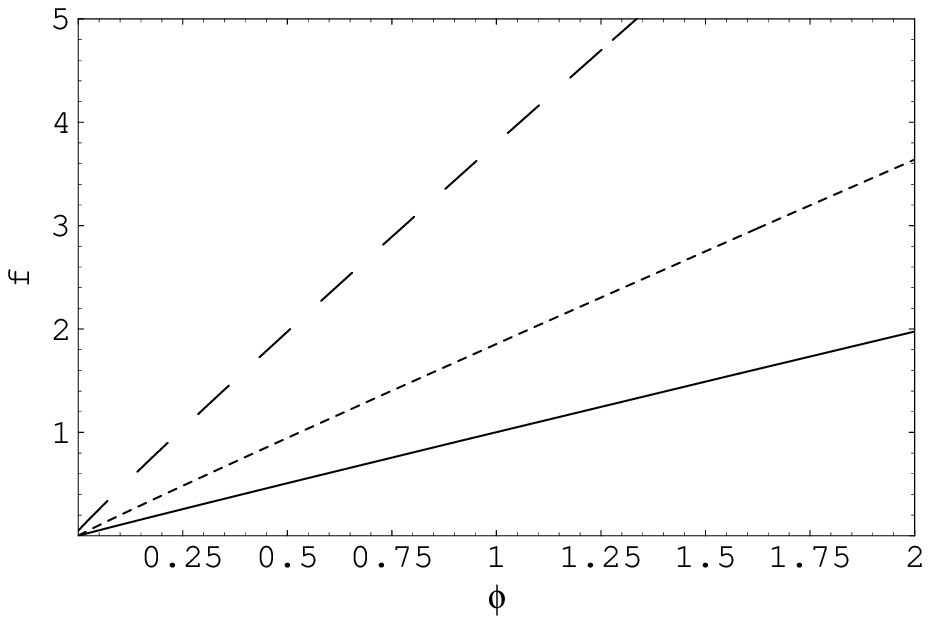}\\
\vspace{1mm}
~~~~~Fig.13~~~~~~~~~~~~~~~~~~~~~~~~~~~~~~~~~~~~~~~~~~~~~~~~~Fig.14\\

\vspace{6mm} ~~~~~~~~~~~~Fig. 13 shows the evolution of the
potential $V$ with chameleon scalar field $\phi$ and Fig. 14 shows
evolution of the analytic function $f$ with chameleon scalar field $\phi$. We have taken $\eta=1/3; c=0.8, 0.5, 0.3$.\\

\vspace{6mm}

\end{figure}

It is observed in the figures 11 and 12 that the chameleon scalar
field $\phi$ and potential $V$ are increasing with cosmic time in
presence of GRDE in emergent universe scenario. In figures 13 and
14 we have plotted $V$ and $f$ both of which are functions of the
chameleon scalar field $\phi$. It is observed from the above
figure that both are increasing functions of $\phi$ in presence of
GRDE in emergent universe scenario. However, $V$ has a sharper
increase than $f$ with increase in the chameleon scalar field
$\phi$.
\\\\

\subsection{\bf\large{Concluding remarks}}
In this paper, generalized Ricci dark energy model, proposed by Xu
et al [27], is presented, where the energy energy density is given
by $\rho_{GR}=3c^{2}M_{pl}^{2}~g\left(\frac{H^{2}}{R}\right)R$.
This generalized form is considered in the emergent universe
scenario. At the beginning of the study, we considered the Ricci
dark energy with energy density $\rho_{R}=3
c^{2}\left(\dot{H}+2H^{2}+\frac{k}{a^{2}}\right)$ in the emergent
universe scenario. The most powerful quantity of dark energy is
its equation-of-state $w$. If we restrict ourselves in
four-dimensional Einstein's gravity, nearly all dark energy models
can be classified by the behaviors of equations of state as
following [35]: (i) for cosmological constant $w=-1$, (ii) for
quintessence $w>-1$, (iii) for phantom $w<-1$ (iv) for quintom $w$
crosses the boundary of $-1$. To discern the behavior of the Ricci
dark energy model we considered in this work, we computed the
equation of state parameter $w$ under different situations. In the
case of ordinary Ricci dark energy we found (fig. 1) that $w$ is
below $-1$ at early stages of the universe and at later stages it
is tending to $-1$. However, it is not crossing the boundary of
$-1$. This means that under emergent universe scenario the
equation of state parameter of Ricci dark energy behaves like
phantom. This observation is in contrast with the observation of
reference [16], where the authors found the crossing of $-1$
boundary by $w$. However, in this work, we have considered
emergent scenario that was not there in [16]. The negative
pressure $p$ is gradually decaying (fig. 2) and dark energy
density is gradually increasing with cosmic time $t$ in this
situation. Next we considered the generalized Ricci dark energy in
the emergent universe scenario and plotted the deceleration
parameter $q$ and fractional energy density $\Omega$ against
cosmic time $t$ for $c=0.8$, $0.5$ and $0.3$. We observed that $q$
is staying a negative level throughout the evolution of the
universe (fig. 4). This is consistent with the emergent (i.e. ever
accelerating) universe. The fractional energy density (fig. 5) is
plotted against cosmic time and it is found that the fractional
energy density is increasing and gradually tending to $1$. This
observation is consistent with the dark energy dominated universe.
Simultaneously we have observed the behaviour of generalized
Holographic dark energy proposed in [27] and observed similar
behaviours in the deceleration parameter and fractional energy
density (figs. 7 and 8). Also, we derived the equation of state
parameters and for generalized Ricci (fig. 6) as well as
generalized holographic dark energy (fig. 9) and we found that in
both of the cases they are are behaving like phantom (i.e.
$w<-1$). Finally we considered the chameleon field and examined
the behaviours of the chameleon scalar field, potential and the
associated function $f$ in presence of generalized Ricci dark
energy in emergent universe scenario for $c=0.8$, $0.5$, $0.3$ and
$\eta=1/3$. In fig. 10 we plotted the function $f$ and observed
its increasing behaviour with cosmic time. The chameleon scalar
field and potentials, plotted in figs. 11 and 12 exhibit
increasing pattern with cosmic time. It was further noted from
figs. 13 and 14 that among $V$ and $f$, the potential $V$ has a
sharper increase with the chameleon scalar field than that of $f$
in the situation under consideration.\\

In summary, it may be noted that the equation of state parameter
for generalized Ricci dark energy is always exhibiting phantom
like behavior under emergent scenario of the universe. It is
already observed from the behavior of the deceleration parameter
that the universe is highly accelerated under the emergent
scenario. Because of this highly accelerated expansion, the
equation of state parameter is remaining at the phantom phase and
is not crossing the $-1$ boundary.
\\\\

{\bf Acknowledgements:}\\\\
The authors sincerely acknowledge the warm hospitality provided by
Inter University Centre for Astronomy and Astrophysics, Pune,
India, where the work was carried out during a scientific visit in
January, 2011. The authors acknowledge the constructive comments
from the reviewers that helped them to enhance the quality of the
manuscript.
\\\\

{\bf References:}\\\\

$[1]$ N.A. Bachall, J.P. Ostriker, S. Perlmutter, P.J. Steinhardt
(1999) \emph{Science} \textbf{284}, 1481;  S. J. Perlmutter et
al.(1999) \emph{Astrophys. J. }\textbf{517}, 565; E. V. Linder
(2003)\emph{ Phys.Rev.Lett.} \textbf{90}, 091301; T. Padmanabhan
(2002)\emph{ Phys.Rev.D}\textbf{66}, 021301.
\\\\
$[2]$ D. N. Spergel et al. [WMAP Collaboration](2003)
\emph{Astrophys. J. Suppl.} \textbf{148}, 175.\\\\
$[3]$ D. J. Eisenstein et al. [SDSS Collaboration](2005)\emph{
Astrophys.J.} \textbf{633}, 560.\\\\
$[4]$ V. Sahni and A. Starobinsty (2006) \emph{International
Journal of Modern Physics D} \textbf{15}, 2105; E. J. Copeland, M.
Sami and S. Tsujikawa (2006) \emph{International Journal of Modern
Physics D} \textbf{15}, 1753; T. Padmanabhan (2005) \emph{Current
Science} \textbf{88}, 1057; T. Padmanabhan (2003) \emph{Physics
Reports} \textbf{380}, 235; D. Huterer and M. S. Turner (2001),
\emph{Phys. Rev. D} \textbf{64}, 123527; V. Sahni (2005)
\emph{Physics of the Early Universe: Lecture Notes in Physics
(Springer)} 653, 141; M. Ishak (2007)\emph{ Foundations of
Physics} \textbf{37}, 1470.\\\\
$[5]$ T. Padmanabhan (2003) \emph{Physics Reports} \textbf{380},
235; S. Weinberg (1989)\emph{Reviews of Modern Physics}
\textbf{61}, 1.\\\\
$[6]$ R.R. Caldwell, R. Dave and P.J. Steinhardt (1998)
\emph{Phys. Rev. Lett.} \textbf{80}, 1582; I. Zlatev, L. Wang, P.
J. Steinhardt (1999) \emph{Phys. Rev. Lett.} \textbf{82}, 896.
\\\\
$[7]$ I. Zlatev, L. M. Wang, and P. J. Steinhardt (1999)\emph{
Phys. Rev. Lett.} \textbf{82}, 896 ; X. Zhang (2005) \emph{Mod.
Phys. Lett. A} \textbf{20}, 2575.
\\\\
$[8]$ R. R. Caldwell (2002) \emph{Phys. Lett. B} \textbf{545}, 23;
R. R. Caldwell, M. Kamionkowski, N. N. Weinberg (2003)\emph{ Phys.
Rev. Lett.} \textbf{91}, 071301.
\\\\
$[9]$ T. Qiu (2010) \emph{Modern Physics Letters A} \textbf{25},
909; Y-F. Cai, E. N. Saridakis, M. R. Setare, J-Q. Xia (2010)
\emph{Physics Reports} \textbf{493}, 1; M. R. Setare, E. N.
Saridakis (2009) \emph{Phys. Rev. D} \textbf{79}, 043005.
\\\\
$[10]$ S. D. Hsu (2004) \emph{Phys. Lett. B} \textbf{594}, 1; A.
G. Cohen, D. B. Kaplan, and A. E. Nelson (1999)\emph{ Phys. Rev.
Lett.} \textbf{82}, 4971; M. Li (2004) \emph{Phys. Lett. B}
\textbf{603}, 1; L. N. Granda, A. Oliveros (2008) \emph{Phys.
Lett. B} \textbf{669}, 275.\\\\
$[11]$ X. Xhang (2009) \emph{Phys. Rev. D} \textbf{79}, 103509.\\\\
$[12]$ P.Kovtun, D.T.Son, A.O.Starinets (2005) \emph{Phys. Rev. Lett.} \textbf{94}, 111601.\\\\
$[13]$ C-J. Feng (2009a)\emph{Physics Letters B} \textbf{672}, 94.\\\\
$[14]$ M. Li (2004) \emph{Phys. Lett. B} \textbf{603}, 1; X. Xhang (2009) \emph{Phys. Rev. D} \textbf{79}, 103509.\\\\
$[15]$ A.G. Cohen, D.B. Kaplan, A.E. Nelson (1999) \emph{Phys.
Rev. Lett.} \textbf{82},
4971 \\\\
$[16]$ C. Gao, F. Wu, X. Chen and Y-G Shen  (2009) {\it Phys. Rev.
D} {\bf 79}, 043511.\\\\
$[17]$ M. Suwa and T. Nihei (2010) \emph{Phys. Rev. D}
\textbf{81},
023519.\\\\
$[18]$ C-J. Feng and X-Z. Li (2009) \emph{Physics Letters B}
\textbf{680}, 355.\\\\
$[19]$ S. Lepe and F. Peña (2010)\emph{ Eur. Phys. J. C}
\textbf{69},
575.\\\\
$[20]$ C-J. Feng (2009b)\emph{ Physics Letters B} \textbf{676}, 168.\\\\
$[21]$ C-J. Feng (2008)\emph{ Physics Letters B} \textbf{670}, 231.\\\\
$[22]$ G. F. R. Ellis and R. Maartens (2004) {\it Class. Quantum
Grav.} {\bf 21}, 223; G. F. R. Ellis, J. Murugan and C. G. Tsagas,
{\it Class. Quantum Grav.} {\bf 21} 233 (2004).\\\\
$[23]$ S. Mukherjee, B. C. Paul, N. K. Dadhich, S. D. Maharaj and
A. Beesham (2006) {\it Class. Quantum Grav.} {\bf 23} 6927.\\\\
$[24]$ S. Mukerji and S. Chakraborty (2011) {\it Astrophys Space
Sci} {\bf 331}, 665.\\\\
$[25]$ U. Debnath (2008), {\it Class. Quantum Grav.} {\bf 25} 205019.\\\\
$[26]$ S. del Campo et al (2007) {\it JCAP} {\bf 11} 030.\\\\
$[27]$ L. Xu, J. Lu and W. Li (2009) {\it Eur. Phys. J. C} {\bf
64},
89.\\\\
$[28]$ S. Nojiri and S. D. Odintsov (2004), {\it Modern Physics Letters A} {\bf 17}, 1273.\\\\
$[29]$ J. Khoury and A. Weltman (2004), {\emph Phys. Rev. D} {\bf
69},
044026\\\\
$[30]$ N. Banerjee, S. Das, K. Ganguly (2010), {\it Pramana} {\bf
74} L481; S. Chakraborty and U. Debnath (2010) \emph{International
Journal of Modern Physics A} \textbf{25},
4691.\\\\
$[31]$ L. Xu and Y. Wang (2010) \emph{JCAP} \textbf{06}, 002,
doi:10.1088/1475-7516/2010/06/002.\\\\
$[32]$ S. Chattopadhyay and U. Debnath \emph{Int J Theor Phys}
\textbf{50},
315.\\\\
$[33]$ L. Xu, W. Li and J. Lu (2009) \emph{Modern Physics Letters
A} \textbf{24},
1355.\\\\
$[34]$ P. B. Khatua and U. Debnath (2010), {\it Astrophys. Space
Sc.}
{\bf 326} 60.\\\\
$[35]$ Y-F. Cai, E. N. Saridakis, M. R. Setare, J-Q. Xia (2010),
{\it Physics Reports} {\bf 493 } 1.
\\\\
\end{document}